# A Novel Supernova Detector*

David B. Cline

*University of California Los Angeles, Dept. of Physics & Astronomy, Box 951547, Los Angeles, CA 90095-1547 USA*

**Abstract.** We discuss the prospects for detecting $\nu_{\mu,\tau}$ and $\nu_\tau$ neutrinos from Type II supernovas using the novel detector at the Supernova Burst Observatory (SNBO) or OMNIS that is being designed for an underground laboratory in the USA. This detector would collect ~2000 flavor selected events from a Galactic supernova and could probe neutrino mass down to a few eV, as well as the dynamics of the supernova process. We believe this is essential to further our understanding of the neutrino section of elementary particle physics.

## INTRODUCTION

The issue of whether or not neutrinos have masses is important for astrophysics and cosmology. Astrophysical considerations may represent the best hope for determining neutrino masses and mixings. In this paper, we examine how proposed neutral-current-based, supernova neutrino-burst detectors, in conjunction with the next generation water-Čerenkov detectors, could use a galactic supernova event to either measure or place constraints on the $\nu_{\mu,\tau}$ masses in excess of 5 eV[1,2]. Such measurements would have important implications for our understanding of particle physics, cosmology, and the solar neutrino problem and would be complementary to proposed laboratory vacuum-oscillation experiments.

A light neutrino mass between 1 eV and 100 eV would be highly significant for cosmology. In fact, if a neutrino contributes a fraction $\Omega_\nu$ of the closure density of the Universe, it must have a mass $m_\nu \approx 92\,\Omega_\nu\,h^2$ eV, where $h$ is the Hubble parameter in units of 100 km s$^{-1}$ Mpc$^{-1}$. Reasonable ranges for $\Omega_\nu$ and $h$ then give 1 eV to 30 eV as a cosmologically significant range. A neutrino with a mass in the higher end of this range (*i.e.*, $10 \le m_\nu \le 30$ eV) could contribute significantly to the closure density of the Universe. The cosmic background explorer (COBE) observation of anistropy in the microwave background, combined with observations at smaller scales, and the distribution of galaxy streaming velocities, have been interpreted as implying that there are two components of dark matter: hot ($\Omega_{HDM} \sim 0.3$) and cold ($\Omega_{CDM} \sim 0.6$). The hot dark matter (HDM) component could be provided by a neutrino with a mass of about 7 eV.[3-5]

## MEASURING THE NEUTRINO MASS BY TIME OF FLIGHT

Perhaps the most straightforward and obvious nature of a massive neutrino would come from the lengthening in flight time from a distant supernova. For example, the flight time difference between $\nu_\tau$ and $\nu_e$ ($\bar{\nu}_e$) in seconds is

$$\Delta t = 5.14 \times 10^{-2} R_{kpc} \left[ \left( \frac{m_{\nu_e}}{E_{\nu_e}} \right)^{-2} - \left( \frac{m_{\nu_x}}{E_{\nu_x}} \right)^{-2} \right] s \quad ,$$

where $E_{\nu_x}$ is the neutrino energy in MeV, $m_{\nu_x}$ is in eV, and $R_{kpc}$ is the distance to the supernova in units of 10 kpc. A finite neutrino mass would alter the neutrino spectra in characteristic ways that could result in broadening and flattening of the observed signal.

Some arguments, which arose during this meeting, for detecting the neutrinos are given in Table 1. Event rates for various detectors for a galactic supernova are given in Table 2.[7] We believe the detection of these supernova neutrino signals will be essential to our understanding of the neutrino sector.

Thus, neutrino masses might be obtained by comparing the observed neutrino signal with the signal expected from supernova models. Since detectors such as Superkamiokande (SK) are relatively insensitive to $\nu_\mu$ and $\nu_\tau$, they are unlikely to measure cosmologically significant neutrino masses for these flavors. One of the neutral-current-based detectors being built at present is the Sudbury Neutrino Observatory (SNO). A general comparison of the methods of measuring neutrino mass is given in Table 3.[1,7] The rate of interaction for the world's detectors is shown in Fig. 1.

---

**Table 1.** Experiments for ν-Mass/Mixing

| Scheme | Tests | | | | | Nucleosynthesis | |
|---|---|---|---|---|---|---|---|
| | $\nu_\odot$ | $\nu_{atmos}$ | LBL | SBL | SN ν's | BBN | SNN |
| I<br>3ν mixing<br>No LSND | Yes<br>$\nu_e \to \nu_{\mu,\tau}$ | Yes<br>$\nu_\mu \to \nu_\tau$ | Yes<br>$\nu_\mu \to \nu_\tau$ | No<br>$\nu_\tau \to \nu_e$<br>$\nu_\mu \to \nu_e$<br>τ appearance? | √ | OK<br><br>$\nu_\mu \to \nu_e$ | OK |
| II<br>4ν mixing<br>$\nu_\mu \to \nu_\tau$<br>(Doublet)<br>$\nu_\odot \to \nu_s$<br>LSND | Yes<br>$\nu_e \to \nu_s$<br>(No extra<br>N.C. signal) | Yes<br>$\nu_\mu \to \nu_\tau$ | Yes<br>$\nu_\mu \to \nu_\tau$ | No?<br>$\nu_e \to \nu_\tau$<br>τ appearance?<br>$\nu_e \to \nu_\tau$?<br>$\nu_e \to \nu_\mu$? | √<br>Hot<br>?<br>Maybe!<br>Maybe! | D/N<br>$\nu_e$-spectrum<br>? | ??<br>Good or bad<br>?? |
| III<br>4ν mixing<br>$\nu_\mu \to \nu_s$<br>Doublet<br>No LSND | Yes<br>$\nu_e \to \nu_\mu$ | Yes<br>$\nu_\mu \to \nu_s$ | Yes<br>$\nu_\mu \to \nu_s$ | No<br>$m_{\nu\tau}$<br>$\nu_\mu \to \nu_\tau$?<br>? | √ | ?<br><br>T of F | r-process<br>constraint |

**Table 2.** Requirements of a Supernova Observatory

- Life of Observatory ≥ rate (yr) for SNII on Milky Way Galaxy ≥ 20 – 40 yr
- Event Rate:     ~ 5 – 10 K     $\bar{\nu}_e + P \to e^+ + n$
               ~ Few K     $\nu_x + N \to \nu_x + N$
                            $\nu_x = \nu_\mu + \nu_\tau$
- To: ∘Fit model of SNII process
      ∘Extract a neutrino mass or neutrino oscillation
      ∘Learn about SNII explosion process

**Table 3.** Methods for Measuring Neutrino Mass

1. Time of flight from an SNII:
$$\Delta t = \frac{1}{2}\left[\left(\frac{M\nu_1}{E_1}\right)^2 - \left(\frac{M\nu_2}{E_1}\right)^2\right] D \text{ sec}$$
Only active neutrinos can separate $M\nu_1, M\nu_2$ by varying $E_1, E_2$.

2. Neutrino oscillation: $P = \sin^2 2\theta \sin^2\left[1.27\left(\frac{L}{E}\right)(M_1^2 - M_2^2)\right]$
Can have sterile neutrinos.

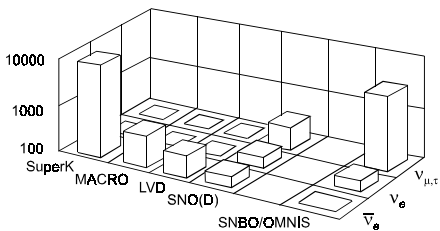

**FIGURE 1.** Comparison of world detectors (event numbers for supernovae at 8 kpc).

## DIFFERENT TYPES OF SN NEUTRINO DETECTORS AND NEUTRINO MASSES

Recently there has been real progress in SN simulations giving an explosion. These calculations give interesting predictions for the neutrino spectra. Detectors like the SK and SNBO/OMNIS may be able to detect such effects, however the SNBO/OMNIS detector may be of crucial importance for this study. Using these various detectors, it should be possible to detect a finite neutrino mass.[1] The characteristics of this detector are listed in Table 4.

In this analysis, we have assumed the existence of a very massive neutral-current detector (SNBO/OMNIS), which we discuss next. By using these different detectors it will be possible to measure the μ or τ neutrino masses, as shown in Table 3, which could determine a mass to ~ 10 eV. To go to lower mass, we need to use the possible fine structure in the burst; we have shown that it may be possible to reach ~ 3 eV with very large detectors in this case.[7] The detection of two-neutron final states, as illustrated in Fig. 2 would be useful for Pb detection.[6]

**Table 4.** Properties of the (Proposed) OMNIS/SNBO Detector

| Targets: | NaCℓ (WIPP site)<br>Fe and Pb (Soudan and Boulby sites) |
|---|---|
| Mass of Detectors: | WIPP site ≥ 200 ton<br>Soudan/Boulby sites ≥ 200 ton |
| Types of Detectors: | Gd in liquid scintillator<br>$^6$Li loaded in the plastic scintillators that are read out by scintillating-fiber–PMT system |

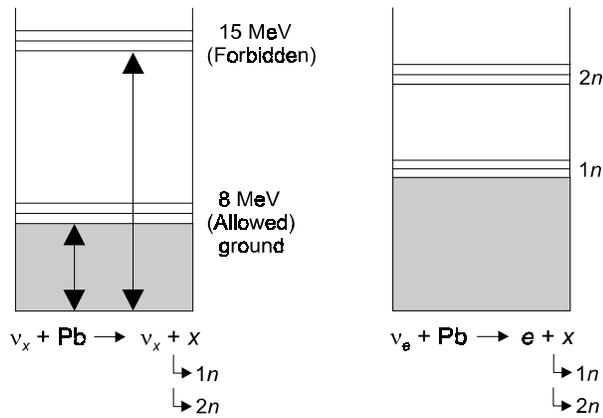

**FIGURE** 2. The two-neutron signal for $\nu_{\tau,\mu} \to \nu_e$. The $2n/1n$ signal is much larger for $\nu_e$ interactions. This is a signature for $\nu_{\tau,\mu} \to \nu_e$ in the SNII environment.

## THE PROPOSED SUPERNOVA BURST OBSERVATORY (SNBO/OMNIS)

The major problem of supernova detection is the uncertain period of time between such processes in this Galaxy. In addition, complimentary detectors should be active when the supernova goes off in order to gain the maximum amount of information possible about the explosion process and neutrino properties. In Table 2, we list some of the requirements of such an ideal supernova observatory.

Lacking an ideal observatory, a group of us have been studying a very large detector, SNBO/OMNIS.[1] Table 4 gives some of the guidelines for this detector.[7] We have located a possible site for the observatory near Carlsbad, NM, which is the WIPP site (shown in Fig. 3). We have studied the radioactive background at this site (measured by the OSU–UCLA group) and find it acceptable for a galactic supernova detector. We find less than one neutron per hour detected in a 6-ft $BF_3$ counter. This leads to the expectation that the background for a galactic supernova is much smaller than the signal at this site. A schematic of the SNBO/OMNIS detector is shown in Fig. 4.

## SEARCH FOR THE INTEGRATED FLUX OF SUPERNOVA NEUTRINOS

Another kind of relic neutrinos are the neutrinos that arise from the integrated flux from all past type-II supernovae. Figure 5 shows a schematic of these (and other) fluxes.[7] These fluxes could be modified by transmission through the SNII environment, as discussed recently.[7]

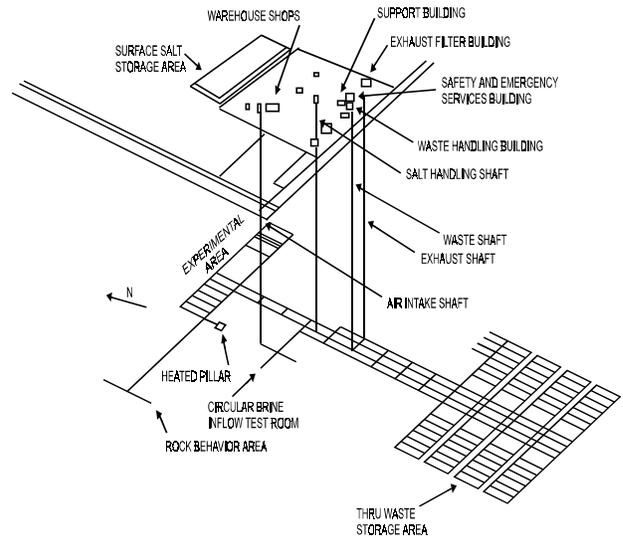

**FIGURE** 3. Illustrative arrangement of detector modules along rock tunnel; also shown is the use of supplementary iron or lead for gamma shielding or as an alternative neutrino target.

The detection of $\bar{\nu}_e$ from the relic supernovae may someday be accomplished by the SK detector. It would be as interesting to detect $\nu_e$ with an ICARUS detector, as illustrated in Table 5. High-energy $\nu_e$ would come from $\nu_{\mu,\tau} \to \nu_e$ in the supernova.[7] A window of detection occurs between the upper solar neutrino energy and the atmospheric neutrinos, as first proposed by D. Cline and reported in the first ICARUS proposal (1983-1985). The ideal detector to observe this is a large ICARUS liquid-argon detector.[7]

## ACKNOWLEDGMENTS

I wish to thank G. Fuller, D. Boyd, K. Lee, and P. F. Smith for discussions.

## REFERENCES

1. Cline, D. B., Fuller, G. M., Hong, W.P., Meyer, B., and J. Wilson, *Phys. Rev.* **D50**, 720 (1994).
2. Presentations to this workshop were made by members of the SNO, SK, AMANDA, SNBO, and KARMAN groups and are to be published in the proceedings (American Institute of Physics, 2000).
3. Smooth, G. F., *et al.*, *Astrophys. J. Lett.* **396**, L1 (1992).
4. Schaefer, R. K., and Shafi, Q.,*Nature* **359**, 199 (1992).
5. Bond, J. R., Efstathiou, G. E., and Silk, J., *Phys. Rev. Lett.* **45**, 1980-1983 (1980).
6. G. Fuller, UCSD, private communication (1998).
7. Cline, D. B. "Search for Relic Neutrinos and Supernova Bursts," in *Proceedings, Eighth Intl. Wksp. on Neutrino Telescopes* (Venice, Feb. 23-26, 1999), ed. M. Baldo Ceolin, 1999, Vol. II, pp. 309-320.

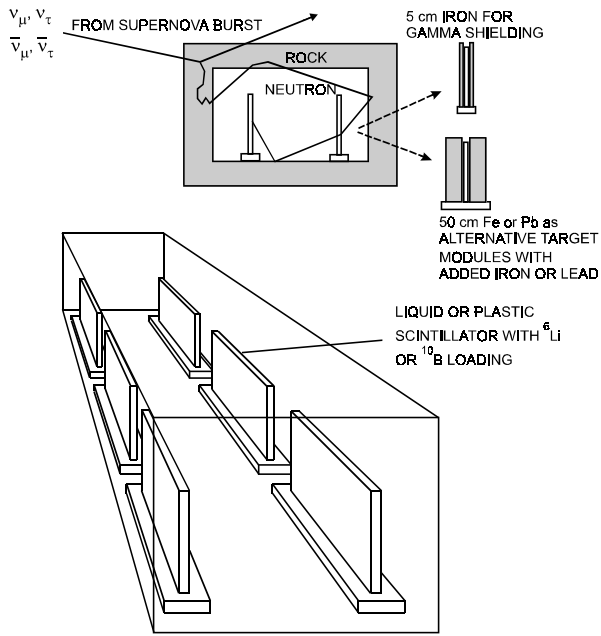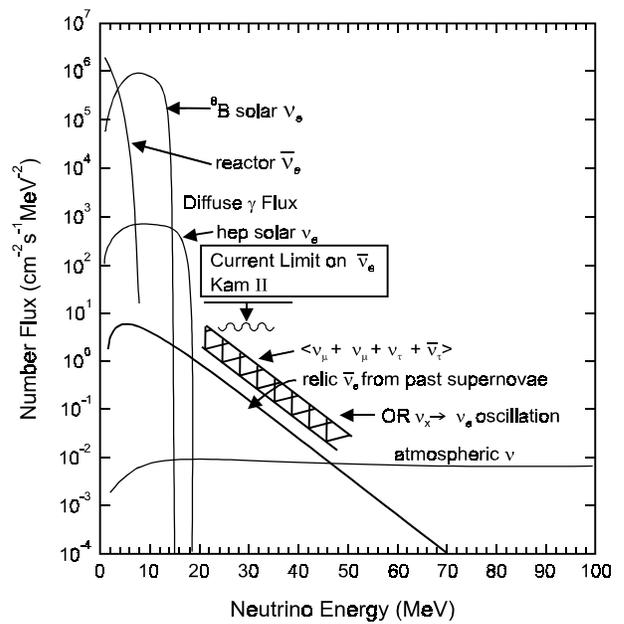

**FIGURE** 4.  Isometric view of the surface and underground (looking toward the Northeast) of the WIPP site near Carlsbad, NM.

**FIGURE** 5.  Relic neutrinos from past supernova. Note: $v_x \to v_e$ in the supernova can boost the energy of the $v_e$ if we find $<E v_e> > <E \bar{v}_e>$. This will be a signal for neutrino oscillation in supernovae! and measure $\sin^2 \theta x_e$.

**Table 5.**  Detection of $v/\bar{v}_e$ Relic Neutrino Flux from Time Integrated SNII
___________________________________________________________________
1. Relic $v/\bar{v}_e$ from all SNII back to $Z \sim 5$:  $<E_v> \sim 1/(1+Z)<E_v>$
2. Detection would give integrated SNII rate from Universe
   – Window of detetion [D. Cline, ICARUS proposal, 1984]
3. Neutrino oscillations in SNII would give $v_x \to v_e$ with higher energy than $\bar{v}_e$
4. Detect $\bar{v}_e$  with SK or ICARUS.  Attempt to detect $v_x/v_e$ detection.
___________________________________________________________________